\pdfoutput=1
\documentclass[11pt]{article}

\usepackage[final]{acl}

\usepackage{times}
\usepackage{latexsym}

\usepackage[T1]{fontenc}
\usepackage[utf8]{inputenc}

\usepackage{microtype}

\usepackage{inconsolata}

\usepackage{graphicx}

\usepackage{amssymb}
\usepackage{subfig}
\usepackage{amsmath}
\usepackage{multirow}

\title{DiffRhythm: Blazingly Fast and Embarrassingly Simple\\
End-to-End Full-Length Song Generation with Latent Diffusion \\
}

\author{
  Ziqian Ning\textsuperscript{1},
  Huakang Chen\textsuperscript{1},
  Yuepeng Jiang\textsuperscript{1},
  Chunbo Hao\textsuperscript{1},\\
  \textbf{Guobin Ma\textsuperscript{1}},
  \textbf{Shuai Wang\textsuperscript{2}},
  \textbf{Jixun Yao\textsuperscript{1}},
  \textbf{Lei Xie\textsuperscript{1}}\\
  \textsuperscript{1}Northwestern Polytechnical University\\
  \textsuperscript{2}Shenzhen Research Institute of Big Data, \\The Chinese University of Hong Kong, Shenzhen (CUHK-Shenzhen), China\\
  }

\begin{document}
\maketitle
\begin{abstract}
Recent advancements in music generation have garnered significant attention, yet existing approaches face critical limitations. Some current generative models can only synthesize either the vocal track or the accompaniment track. While some models can generate combined vocal and accompaniment, they typically rely on meticulously designed multi-stage cascading architectures and intricate data pipelines, hindering scalability. Additionally, most systems are restricted to generating short musical segments rather than full-length songs. Furthermore, widely used language model-based methods suffer from slow inference speeds.
To address these challenges, we propose DiffRhythm, the first latent diffusion-based song generation model capable of synthesizing complete songs with both vocal and accompaniment for durations of up to 4m45s in only ten seconds, maintaining high musicality and intelligibility. Despite its remarkable capabilities, DiffRhythm is designed to be simple and elegant: it eliminates the need for complex data preparation, employs a straightforward model structure, and requires only lyrics and a style prompt during inference. Additionally, its non-autoregressive structure ensures fast inference speeds. This simplicity guarantees the scalability of DiffRhythm. Moreover, we release the complete training code along with the pre-trained model on large-scale data to promote reproducibility and further research\footnote{https://nzqian.github.io/DiffRhythm/}%\footnote{https://anonymous.4open.science/w/DiffRhythm-3EBC/}.

%This work bridges the gap between conceptual text-to-music research and practical, high-fidelity song synthesis, offering a robust foundation for future innovations in creative AI-driven music production.

%In this paper, we propose DiffRhythm, the first flow-matching-based song generation model capable of generating full songs up to 3 minutes. With ultra simple model structure and minimal data requirements, DiffRhythm is easy to scale up. Experiments demonstrate that the proposed DiffRhythm achieves exceptional performance in end-to-end song generation, as evidenced by both objective and subjective metrics. Code and pre-trained weights are available in \footnote{repo}

\end{abstract}

\section{Introduction}
Music, as a form of artistic expression, holds profound cultural importance and resonates deeply with human experiences~\cite{music-survey}. The field of music generation has witnessed remarkable advancements in recent years, driven by innovations in deep learning, particularly the deep generative models.

% Researchers have explored a variety of techniques to synthesize music, leading to the development of numerous generative models. 
While these models have shown promise, they often exhibit critical limitations that restrict their practical applicability. Many existing approaches are designed to generate vocal tracks and accompaniment tracks independently, resulting in a disjointed musical experience. For instance, studies such as Melodist~\cite{melodist} and MelodyLM~\cite{melodylm} demonstrate the effectiveness of isolated track generation, yet highlighting the need for more holistic solutions that capture the interplay between vocals and accompaniment.

Currently, there are relatively few studies on end-to-end song generation in the academic field. State-of-the-art platforms like Seed-Music~\cite{seed-music} and Suno\footnote{https://suno.com/} are generally for commercial products and provides no open-source implementation or detailed technical documentation. 

Recent academic work such as SongCreator~\cite{songcreator} and SongEditor~\cite{songeditor} endeavor to create combined vocal and accompaniment outputs; however, these typically rely on complex, multi-stage cascading architectures. This complexity not only complicates design and implementation but also limits scalability, particularly for longer audio synthesis where maintaining consistency is challenging. The ability to generate complete compositions is essential for practical applications in both artistic creation and commercial music production.
Moreover, most existing music generation models follow a language model paradigm~\cite{melodist, melodylm, songeditor, musiclm}, often struggling with slow inference speeds, which hinder real-time applications and user interactivity. 
% As a result, there is a pressing need for innovative frameworks that can streamline the music generation process while enhancing efficiency and output quality.

To address these challenges, we present DiffRhythm, the first full-diffusion-based song generation model that is capable of synthesizing full-length songs comprising both vocal and accompaniment for durations of up to four minutes. DiffRhythm distinguishes itself not only through its ability to maintain high levels of musicality and intelligibility but also through its simple yet effective model architecture and data processing pipeline, designed specifically for scalability. Additionally, our non-autoregressive approach allows for fast generation speeds, significantly improving usability compared to current models. The main contributions of this paper are summarized as follows:
\begin{itemize}
    \item We propose DiffRhythm, the first end-to-end diffusion-based song generation model capable of generating full song with both vocal and accompaniment.
    \item We propose a sentence-level lyrics alignment mechanism for better vocal intelligibility, which tackles ultra-sparse lyrics-vocal alignment with minimal supervision.
    \item We train a Variational Autoencoder (VAE) tailored for high-fidelity music reconstruction,while demonstrating exceptional robustness against MP3 compression artifacts. Moreover, our VAE shares the same latent space with the famous Stable Audio VAE\footnote{https://github.com/Stability-AI/stable-audio-tools\label{stableaudio}}, enabling seamless plug-and-play substitution in existing latent diffusion frameworks.
    \item Our experiments show that despite its simpleness, DiffRhythm achieves excellent performance in song generation. The data processing pipeline, pretrained models trained on large-scale datasets, and the complete training recipe are publicly available.
\end{itemize}

%By releasing the complete training code and model weights, we aim to promote reproducibility and encourage further exploration in this vital area of research. In the subsequent sections, we will delve into the methodology and experiments underpinning DiffRhythm, elucidating its contributions to the evolving landscape of music generation research.

\section{Related Work}

\begin{figure*}[!htbp]
\centering
\includegraphics[scale=0.6]{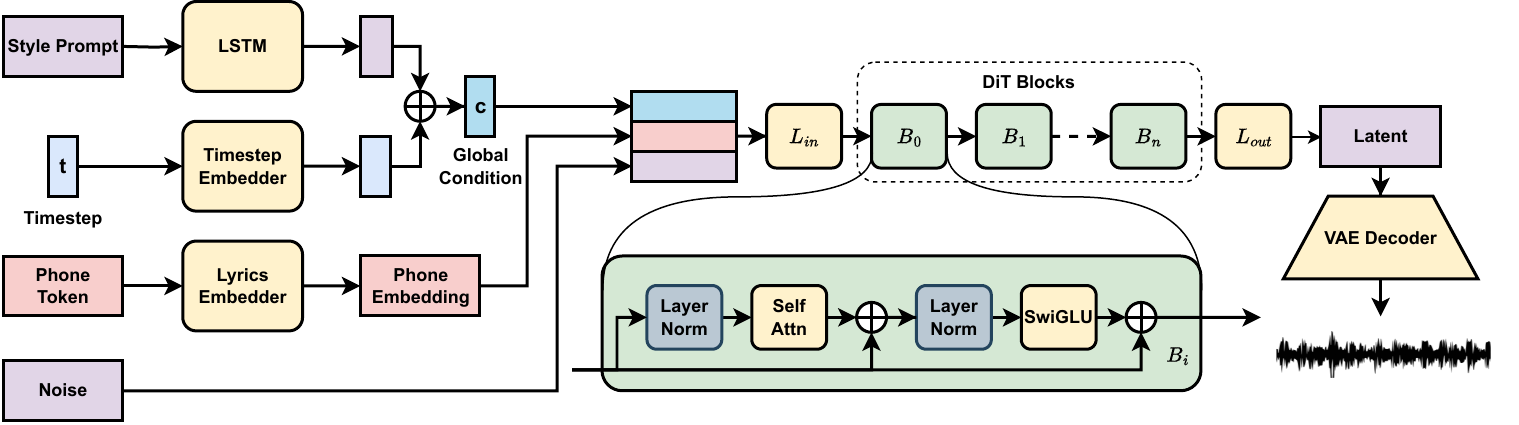}
\caption{Architecture of DiffRhythm. The style and lyrics are used as external control signals, which are preprocessed to get the style embedding and lyrics token, input to DiT to generate latent, and subsequently go through the VAE decoder to generate the audio.}
% \vspace{-8pt}
\label{fig:model}
\end{figure*}
\subsection{Vocal Generation}
Early models for vocal generation, or singing voice generation, focused on synthesizing natural singing voices based on lyrics, musical scores, and corresponding durations. VISinger 2~\cite{visinger2} introduces an end-to-end system utilizing a digital signal processing (DSP) synthesizer to enhance sound quality. StyleSinger~\cite{stylesinger} employ a reference voice clip for timbre and style extraction, enabling style transfer and zero-shot synthesis. PromptSinger~\cite{promptsinger} was the first system to attempt guiding singing voice generation through text descriptions, placing greater emphasis on timbre control. DiffSinger~\cite{diffsinger} addresses the issue of excessive smoothness by implementing a shallow diffusion mechanism. To bridge the gap between realistic music scores and detailed MIDI annotations, RMSSinger~\cite{rmssinger} proposes a word-level modeling approach combined with diffusion-based pitch prediction. MIDI-Voice~\cite{midivoice} incorporates MIDI-based priors for expressive zero-shot generation. VoiceTuner~\cite{voicetuner} advocates a self-supervised pre-training and fine-tuning strategy to mitigate data scarcity, applicable to low-resource SVS tasks. There are also recent models that do not rely on strict music score and duration annotations, such as Freestyler~\cite{freestyler}, which takes lyrics and accompaniment as inputs to generate rapping vocals with strong stylistic and rhythmic alignment with accompanying beats.

\subsection{Music Generation}
Music generation encompasses various tasks, including symbolic music generation, lyrics generation, and accompaniment generation. MuseGAN~\cite{musegan} achieves symbolic music generation through a GAN-based approach. SongMASS~\cite{songmass} designs a method for songwriting that generates lyrics or melodies conditioned on each other, while SongComposer~\cite{songcomposer} proposes a large language model (LLM) for song composition, capable of generating melodies and lyrics with symbolic song representations. DeepRapper~\cite{deeprapper} focuses on rap lyrics generation, which also leverages an LLM to generate lyrics from right to left with rhyme constraints.
Inspired by two-stage modeling in audio generation~\cite{audiolm}, MusicLM~\cite{musiclm} uses a cascade of transformer decoders to sequentially generate semantic and acoustic tokens, based on joint textual-music representations from MuLan~\cite{mulan}. MusicGen~\cite{musicgen} introduces a novel approach with codebook interleaving patterns to generate music codec tokens in a single transformer decoder, which is further combined with stack patterns in ~\citealp{stack-and-delay} to improve generation quality. Additionally, MeLoDy~\cite{melody} presents an LM-guided diffusion model that efficiently generates music audio, and MusicLDM~\cite{musicldm} incorporates beat-tracking information and latent mixup data augmentation to address potential plagiarism issues in music generation.
Several works focus specifically on vocal-to-accompaniment generation, such as SingSong~\cite{singsong}, which generates instrumental music to accompany input vocals, and Melodist~\cite{melodist}, which utilizes a transformer decoder for controllable accompaniment generation.

\subsection{Song Generation}
Song generation models aim to produce natural singing voices accompanied by music. Song generation incorporates elements from both vocal and music generation. A common methodology in song generation employs a two-stage process: initially generating the vocal track from lyrical input, followed by the prediction of accompanying music. Melodist~\cite{melodist} utilizes two autoregressive transformers to sequentially produce vocal and accompaniment codec tokens, conditioned on lyrics, musical scores, and natural language prompts. 
MelodyLM~\cite{melodylm} eliminates the need for music scores in Melodist and instead relies solely on textual descriptions and vocal references. However, given the intricate relationship between vocals and accompaniment, sequential generation may not be optimal. Different from Melodist and MelodyLM, SongCreator~\cite{songcreator} simultaneously generates vocal and accompaniment, while SongEditor~\cite{songeditor} also offering flexible song editing capabilities. It is noteworthy that these models predominantly utilize language model-based architectures.  While effective, their autoregressive nature introduces significant computational overhead and challenges in maintaining consistent style and rhythm over long sequences.

\section{DiffRhythm}

\begin{figure*}[!htbp]
\centering
\includegraphics[scale=0.70]{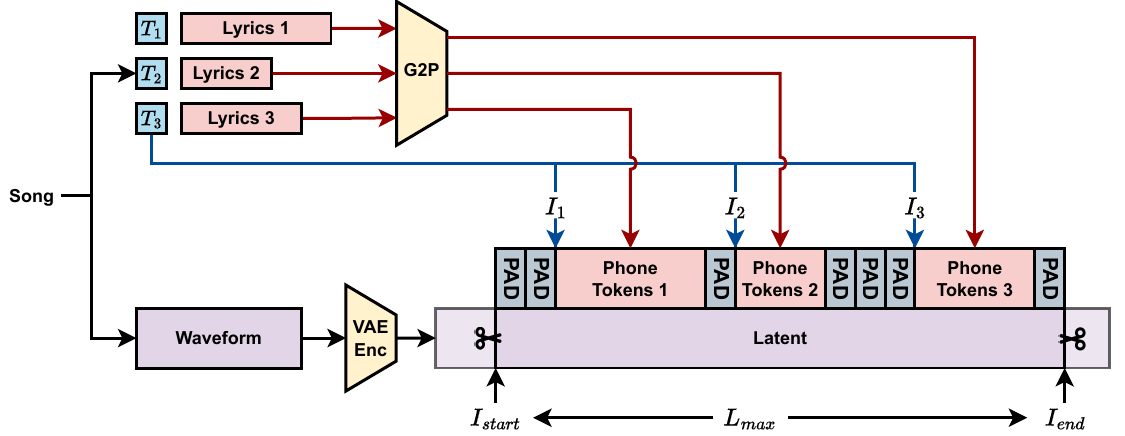}
\caption{The data preprocessing pipeline of DiffRhythm. Lyrics go through G2P and are placed at the positions corresponding to their timestamps}
\label{fig:feat}
%\vspace{-8pt}
\end{figure*}
To address the limitations of existing approaches and overcome the challenges in full-length song generation, we present DiffRhythm - the first full-diffusion-based model specifically designed for end-to-end song generation.
\subsection{Overview}
DiffRhythm produces full-length stereo musical compositions (up to 4m 45s) at 44.1kHz sampling rate, guided by lyrics and style prompts. The architecture consists of two consecutively trained models : 1) A variational autoencoder (VAE) that learns compact latent representations of waveforms while preserving perceptual audio details, effectively resolving the sequence length constraints in raw audio modeling; 2) A diffusion transformer (DiT) operating in the learned latent space that generates songs through iterative denoising. Compared with conventional discrete tokens in LM-based approaches, our continuous latent representation captures richer music details and vocal nuances, enabling high-fidelity audio reconstruction. Meanwhile, the DiT's strong modeling capabilities and the reduced sequence length of continuous VAE latents ensure superior long-term musical structure consistency and vocal intelligibility across full-length songs.

%Furthermore, DiffRhythm eliminates the need for explicit time-aligned lyric-vocal data, a complex and often inaccurate preprocessing step. Instead, to tackle the critical challenge of lyric-vocal alignment in full-song generation, we propose a novel sentence-level alignment mechanism that establishes semantic correspondence between tight lyrical content and sparse singing vocals. 
Furthermore, to tackle the critical challenge of lyric-vocal alignment in full-song generation, we propose a novel sentence-level alignment mechanism to establish semantic correspondence between dense lyrical content and sparse singing vocals.
% MOVE TO ALIGNMENT SUBSECTION: This approach achieves high intelligibility while minimizing the reliance on supervision, effectively reducing the cost of data labeling processing.

\subsection{Variational Autoencoder}
To lower the computational demands of training the diffusion model towards long-form high-quality song generation, we first train an autoencoding model which learns a latent space that is perceptually equivalent to the audio space, but offers significantly reduced computational complexity. 

\textbf{Model Backbone}\quad
The backbone of the autoencoder is fully-convolutional that allows the compression and reconstruction of full-songs with arbitrary-length. The encoder and decoder structures are taken from Stable Audio 2~\cite{stable-audio}. Given a raw stereo waveform $y \in \mathbb{R}^{T\times2}$, the encoder $\mathcal{E}$ encodes $y$ into a latent representation $z = \mathcal{E}(y)$, the decoder $\mathcal{D}$ reconstructs the song from the latent, giving $\hat{y} = \mathcal{D}(z) = \mathcal{D}(\mathcal{E}(y))$, where $z \in R^{L\times c}$. The encoder downsamples the audio by a factor $f = T/L$. 

\textbf{Training Objectives}\quad
The VAE is optimized through a composite loss function integrating spectral reconstruction and adversarial training components. The primary training objective combines a multi-resolution STFT loss~\cite{stftloss} with perceptual weighting, specifically designed for stereo signal processing. To address potential ambiguities in spatial localization, we compute this loss in both mid-side (M/S) decomposition and individual left/right channel domains, with the latter contribution scaled by 0.5 relative to the M/S term. 

Complementing this reconstruction objective, we implement an adversarial training scheme using a convolution-based discriminator~\cite{disc}. While maintaining hyperparameters with Stable Audio~\cite{stable-audio1}, the discriminator features substantially expanded channel dimensions, resulting in approximately quadrupled parameter count compared to the original implementation. This enhancement aims to improve the model's capacity for capturing high-frequency audio details through more discriminative feature learning.

\textbf{Lossy-to-Lossless Reconstruction}\quad
Considering that a large amount of song data exists in compressed MP3 format, where high-frequency components are compromised during compression, we employ data augmentation to equip the VAE with restoration capabilities. Specifically, the VAE is trained exclusively on lossless FLAC-format data, where the input undergoes MP3 compression while the reconstruction target remains the original lossless data. Through this lossy-to-lossless reconstruction process, the VAE learns to decode latent representations derived from lossy-compressed data back into lossless audio signals.

\textbf{Latent Truncation for Training}\quad
As illustrated in Figure~\ref{fig:feat}, for diffusion training, we randomly sample a starting frame index $I_{start}$ and truncate $z$ from $I_{start}$ to a feature length of $L_{max}$ for batch consistency. Another small segment of latent is also randomly selected and used as the style prompt to provide style information. Specific length configurations are detailed in Section~\ref{sec:exp-setup}.

\subsection{Diffusion Transformer}
With compact latent features extracted by the VAE encoder as intermediate representations, we adopt the widely used diffusion transformer (DiT) for lyrics-to-latent generation. DiT has seen notable success in other modalities~\cite{dit, stable-diffusion3}, and has recently been applied to text-to-speech~\cite{ardit,e2tts, f5tts} and music generation~\cite{stable-audio, fluxmusic, tangoflux}. 

\textbf{Feature Conditioning}\quad
As shown in Figure~\ref{fig:model}, DiT is conditioned by three features: A style prompt for controlling song style, a timestep indicating the current diffusion step, and lyrics for vocal content control. The style prompt goes through a Long Short-Term Memory (LSTM) network, where the final hidden state is extracted as the global style information. This information is then summed with the time-step embedding to form a global condition feature. The phone tokens of the lyrics undergo processing through an embedding layer to produce continuous phoneme embeddings. Following this, latent representations undergo noise addition to get noised latent. These three features are concatenated along the channel dimension to serve as inputs to DiT. The feature extraction process will be detailed in Sec.~\ref{sec:lla}.

\textbf{Model Backbone}\quad
Different from the original DiT implementation~\cite{dit}, DiT in DiffRhythm incorporates stacks of LLaMA decoder layers. Given that LLaMA is widely used in natural language processing (NLP), several readily available acceleration libraries, such as Unsloth\footnote{https://github.com/unslothai/unsloth} and Liger-Kernel\footnote{https://github.com/linkedin/Liger-Kernel}, that can easily achieve more than 25\% training and inference speed-ups relative to the original DiT without any performance degradation through kernel fusion. We employ efficient FlashAttention2~\cite{flashattention2} and gradient checkpointing~\cite{grad-ckpt} to reduce the computational and memory impact of applying a transformer architecture over longer sequences. These techniques are essential for the effective training of models with extensive context lengths.

\begin{figure}[ht]
\centering
\includegraphics[scale=0.5]{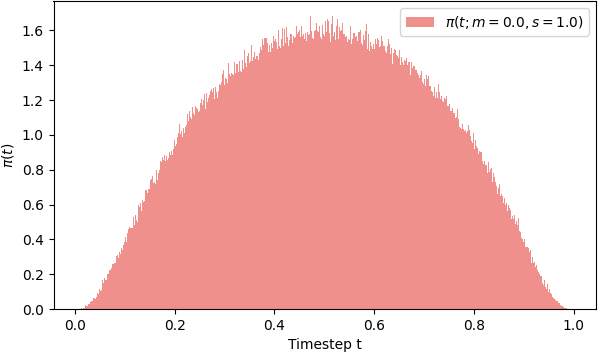}
\caption{Logit-normal timestep distribution.}
\label{fig:logit-norm}
% \vspace{-8pt}
\end{figure}

% \textbf{Timestep Samping}\quad
% The RF loss trains the velocity v uniformly on all timesteps in [0, 1]. Intuitively, however, the resulting velocity prediction target   is more difficult for t in the middle of [0, 1]. One option for a distribution that puts more weight on intermediate steps is the logit-normal distribution. Its density,
% \begin{equation}
% \pi_{ln}(t;m, s) = \frac{1}{s\sqrt{2\pi}}\frac{1}{t(1-t)}\text{exp}(-\frac{(\text{logit}(t)-m)^2}{2s^2})
% \end{equation}
% where $\text{logit}(t) = \text{log}\frac{t}{1-t}$ , has a location parameter, m, and a scale parameter, s. The location parameter enables us to bias the training timesteps towards either data $p_0$ (negative m) or noise $p_1$ (positive m). As shown in Figure 11, the scale parameters controls how wide the distribution is.  In practice, we sample the random variable u from a normal distribution $u \sim \mathcal{N} (u; m, s)$ and map it through the standard logistic function.

%DiffRhythm's streamlined pipeline significantly reduces preprocessing complexity compared to conventional approaches requiring metediously designed model strcture and multi-stage processing (e.g., source separation, structural annotations). As shown in Figure~\ref{fig:feat} framework operates on five minimally processed features:

%\textbf{Style Embedding Extraction:} Free-form style descriptors (e.g., "synthwave retro", "soulful R\&B") are encoded through a frozen T5 text encoder to produce style embeddings $s \in \mathbb{R}^{l_s\times d_s}$.

\textbf{Training Objectives}\quad
Following the conditional flow matching paradigm~\cite{flow-matching}, our model learns a velocity field $v_\theta(z_t, t)$ that transports the noise distribution $p_0(z)$ to the data distribution $p_1(z)$ through the ODE:

\begin{equation}
    \frac{dz_t}{dt} = v_\theta(z_t, t) \quad \text{with} \quad 
    \begin{cases}
        z_0 \sim p_0(z) \\
        z_1 \sim p_1(z)
    \end{cases}
\end{equation}

The training objective minimizes the expected squared error between predicted and target velocity fields:

{\small
\begin{equation}
    \mathcal{L} = \mathbb{E}_{t \sim \pi_{\text{ln}}, z_t \sim p_t(z_t)} \left[\|v_\theta(z_t, t, c) - (z_1 - z_0)\|_2^2\right],
\end{equation}
}where $c$ is the condition, and the timestep sampling distribution $\pi_{\text{ln}}(t;m,s)$ follows the logit-normal density:

% \begin{equation}
% \begin{split}
%     \pi_{\text{ln}}(t;m,s) &= \frac{1}{s\sqrt{2\pi}}\frac{1}{t(1-t)} \\
%     &\qquad \exp\left(-\frac{(\mathrm{logit}(t)-m)^2}{2s^2}\right)
% \end{split}
% \end{equation}
{\small
\begin{equation}
    \pi_{\text{ln}}(t;m,s) = \frac{1}{s\sqrt{2\pi}}\frac{1}{t(1-t)} \exp\left(-\frac{(\mathrm{logit}(t)-m)^2}{2s^2}\right),
\end{equation}
}with $\mathrm{logit}(t) = \log\frac{t}{1-t}$. As discussed in Stable Diffusion 3~\cite{stable-diffusion3}, logit-normal sampling provides adaptive weighting where the scale parameter $s$ controls concentration around mid-point timesteps (challenging prediction regions), while the location parameter $m$ enables bias toward either data ($m<0$) or noise ($m>0$) domains. This allows training to focus more effectively on complex intermediate regions. In practice, we sample $u \sim \mathcal{N}(m, s)$ and map it through the logistic function $t = \sigma(u) = 1/(1+e^{-u})$. Figure~\ref{fig:logit-norm} illustrates the timestep distribution when $m=0$ and $s=1$.

\begin{table*}[ht]
\centering
\renewcommand\arraystretch{1.4}
\caption{Comparative evaluation of waveform reconstruction performance using objective metrics. STOI, PESQ, and MCD scores are reported for both lossless-to-lossless and lossy-to-lossless reconstruction.}
\label{tab:ae}
\resizebox{1.0\linewidth}{!}{
\begin{tabular}{lccccccccc}
\hline
               & \multicolumn{3}{c}{Lossless $\rightarrow$ Lossless}                & \multicolumn{3}{c}{Lossy $\rightarrow$ Lossless}                   &                           &            &                     \\ \cline{2-7}
               & STOI$\uparrow$ & PESQ$\uparrow$ & MCD$\downarrow$ & STOI$\uparrow$ & PESQ$\uparrow$ & MCD$\downarrow$ & Sampling Rate             & Frame Rate & Latent Channels     \\ \hline
Music2Latent   & 0.584          & 1.448          & 8.796           & -              & -              & -               & \multirow{3}{*}{44.1 kHz} & 10 Hz      & \multirow{3}{*}{64} \\
Stable Audio 2 VAE & 0.621          & 1.96           & 8.033           & -              & -              & -               &                           & 21.5 Hz    &                     \\
DiffRhythm VAE       & \textbf{0.646}          & \textbf{2.235}          & \textbf{8.024}           & \textbf{0.639}          & \textbf{2.191}          & \textbf{9.319}           &                           & 21.5 Hz    &                     \\ \hline
\end{tabular}}
%\vspace{-8pt}
\end{table*}

\subsection{Lyrics-to-Latent Alignment} 
\label{sec:lla}
%Song generation demands models capable of producing audible vocal content, which requires establishing accurate lyric-vocal correspondence. This task fundamentally differs from text-to-speech (TTS) systems designed for relatively shorter speech segments (typically <30s) with continuous articulation. Specifically, song generation must address two critical challenges: (1) extremely sparse lyric-vocal correspondence: two vocal segments may be separated by long instrumental spans that break phonetic continuity, and (2) accompaniment interference: The pronunciation of the same word in different songs with different accompaniments gives the model extra alignment difficulty. The naive lyrics conditioning approach, such as cross attention or direct concatenation~\cite{e2tts}, fails to generate vocals with intelligibility. 

%To tackle the ultra-sparse alignment challenge, we propose sentence-level aligning, which only requires minimal supervision to achieve good intelligibility.
%First we create a sequence of equal length to latent, i.e. of length $T_{max}$, filled with $[pad]$ tokens. Subsequently, based on $T_{start}$ and $T_{end}$, active lyrics sentences are determined through their timestamp annotations. Each qualifying lyric sentence undergoes grapheme-to-phoneme (G2P) conversion to obtain phoneme tokens. For the phone token of each sentence, replace the pad token at the corresponding position according to its timestamp. Taking sentence-level aligned phone tokens as input greatly reduces the alignment difficulty of the model and ensures intelligibility of the human voice.

Song generation, which necessitates the creation of intelligible vocal content, presents unique alignment challenges beyond conventional text-to-speech (TTS) task. While TTS models typically handle shorter speech segments (usually less than 30 seconds) with continuous articulation, vocal generation must address two critical alignment problems:

(1) \textit{Discontinuous temporal correspondence}: Vocal segments are often separated by prolonged instrumental intervals, creating phonetic discontinuity that disrupt conventional temporal alignment mechanisms.

(2) \textit{Accompaniment interference}: As we target to simultaneously model voice and accompaniment, the same words, although corresponding to the same pronunciation, have different accompaniment in different songs, which brings more difficulty in aligning.

With conventional text conditioning approaches in diffusion-based TTS models like cross-attention mechanisms or direct feature concatenation~\cite{e2tts, f5tts}, we failed to achieve intelligibility in song generation.

%To address the ultra-sparse alignment challenge, we propose a sentence-level alignment strategy requiring minimal supervisory labels. Our method constructs a latent-aligned phone sequence $\mathbf{P} \in \mathcal{V}^{F_{max}}$ where $\mathcal{V}$ denotes the phoneme vocabulary and $F_{max}$ represents the maximum training sequence length. The sequence is initialized with padding tokens, with each lyric sentence processed through three stages according to its annotated start timestamp $F_{start}$. First, text-based lyrics undergo grapheme-to-phoneme (G2P) conversion to obtain phone tokens. These tokens are then anchored at the computed starting frame index $f_{start} = \lfloor t_{start} \cdot F_s \rfloor$, where $F_s$ denotes the latent frame rate. Subsequent phonemes sequentially occupy consecutive positions in $\mathbf{P}$ starting from $f_{start}$, overwriting the original padding tokens. Taking sentence-level aligned phone sequence as input greatly reduces the alignment difficulty of the model and ensures vocal intelligibility.

It is relatively challenging for the model to tackle both tasks simultaneously. Therefore, we aim to reduce the difficulty of alignment, allowing the model to focus more on the second challenge. To achieve this, we propose a sentence-level alignment paradigm that requires only sentence-start annotations.
Given lyric sentences with timestamp annotations ${ (t_i^{start}, s_i) }_{i=1}^N$, we first convert each lyric sentence $s_i$ into a phoneme sequence $\mathbf{p}_i \in \mathcal{V}^{L_i}$ through grapheme-to-phoneme (G2P) conversion, where $\mathcal{V}$ denotes the phoneme vocabulary and $L_i$ denotes the sequence length of $s_i$.  
Next, we initialize a latent-aligned sequence $\mathbf{P}_i = [\langle \text{pad} \rangle]^{L_{max}}$ with the same length as the latent representation. Then, for each phoneme sequence $\mathbf{p}_i = [p_1, \dots, p_{L_i}]$, we overwrite the corresponding section of $\mathbf{P}_i$ as follows:  $\mathbf{P}_i[f_i^{start} : f_i^{start} + L_i] = \mathbf{p}_i$,  $f_i^{start} = \lfloor t_i^{start} \cdot F_s \rfloor,$ where $F_s$ denotes the latent frame rate. The whole process is detailed in Figure~\ref{fig:feat}.
The proposed approach achieves high intelligibility while minimizing the reliance on supervision, effectively reducing the cost of data labeling processing.
%Crucially, these phoneme tokens are time-anchored to frame indices proportional to their vocal onset timestamps relative to $T_{start}$ , while non-vocal regions (including instrumental intervals and inter-phrase pauses) are populated with [PAD] tokens (zero-initialized vectors). This constructs a sparse lyrics feature $l \in \mathbb{R}^{l\times d}$ strictly aligned with the latent dimension $l$.

\section{Experimental Setup}
\label{sec:exp-setup}

\subsection{Dataset}
DiffRhythm was trained on a comprehensive music dataset comprising approximately 1 million songs (totaling 60,000 hours of audio content) with an average duration of 3.8 minutes per track. The dataset features a multilingual composition ratio of 3:6:1 for Chinese songs, English songs, and instrumental music respectively. To ensure lyrical quality, we implemented a simple rule-based lyrics cleaning pipeline that systematically filters out low-quality lyrics. Subsequently we pre-extract the phoneme tokens from lyrics using MaskGCT~\cite{maskgct} G2P and song latent using the pre-trained VAE for faster training.

For autoencoder evaluation, we selected 10 representative music genres, sampling three tracks per genre to form a 30-song test set. Five non-overlapping 10-second clips were randomly extracted from each track for analysis. To assess the song generation quality, we reserved 30 songs from the training dataset and generated samples using ground-truth lyrics and style prompts as input conditions.

\subsection{Model Configuration}
\textbf{VAE}\quad
Our implementation adapts the pre-trained weights from Stable Audio 2's VAE with 157M parameters, freezing the encoder while training the decoder for 2.5M iterations on a curated dataset of 250k lossless audio samples. The architecture processes 44.1 kHz stereo audio inputs through 5× downsampling blocks achieving a compression factor of $f=2048$, yielding 64-dimensional latent representations at 21.5 Hz frame rate. During training, there was a 1/3 probability of keeping the inputs unchanged and a 2/3 probability of applying MP3 compression, with uniformly randomized VBR\footnote{Variable bit rate, lower values represent higher quality} quality value from 0 to 7. MP3 compression is achieved using pedalboard\footnote{https://github.com/spotify/pedalboard}.

\noindent \textbf{DiT}\quad
Our DiT implementation comprises 16 LLaMA decoder layers\footnote{https://github.com/huggingface/transformers} with 2048-dimensional hidden size and 32-head self-attention mechanisms (64 dimensions per head), totaling 1.1B parameters. We apply independent 20\% dropout to lyrics and style prompts to facilitate classifier-free guidance (CFG)~\cite{cfg}. The diffusion process employs an Euler ODE solver with 32 steps and CFG scale of 4 during inference. Training occurs in two phases: initial base model training with $L_{max}=2048$ ($\approx$ 95s), followed by fine-tuning to $L_{max}=6144$ ($\approx$ 4m45s).

Both models were trained using AdamW optimizer with $\beta_1=0.9$ and $\beta_2=0.95$. The learning rate was set to $1 \times 10^{-4}$ with exponential ramp-up and decay. To ensure model stability and performance, we maintain a secondary copy of the model weights, updated every 100 training batches through an exponential moving average (EMA) with a decay rate of 0.99, following the approach outlined by~\cite{dit}. All models were trained on 8x Huawei Ascend 910B with fp16 mixed-precision.
\begin{figure}[ht!]
    \centering
    % 第一个子图
    \subfloat[GT (Lossless)]{%
        \includegraphics[width=0.23\textwidth]{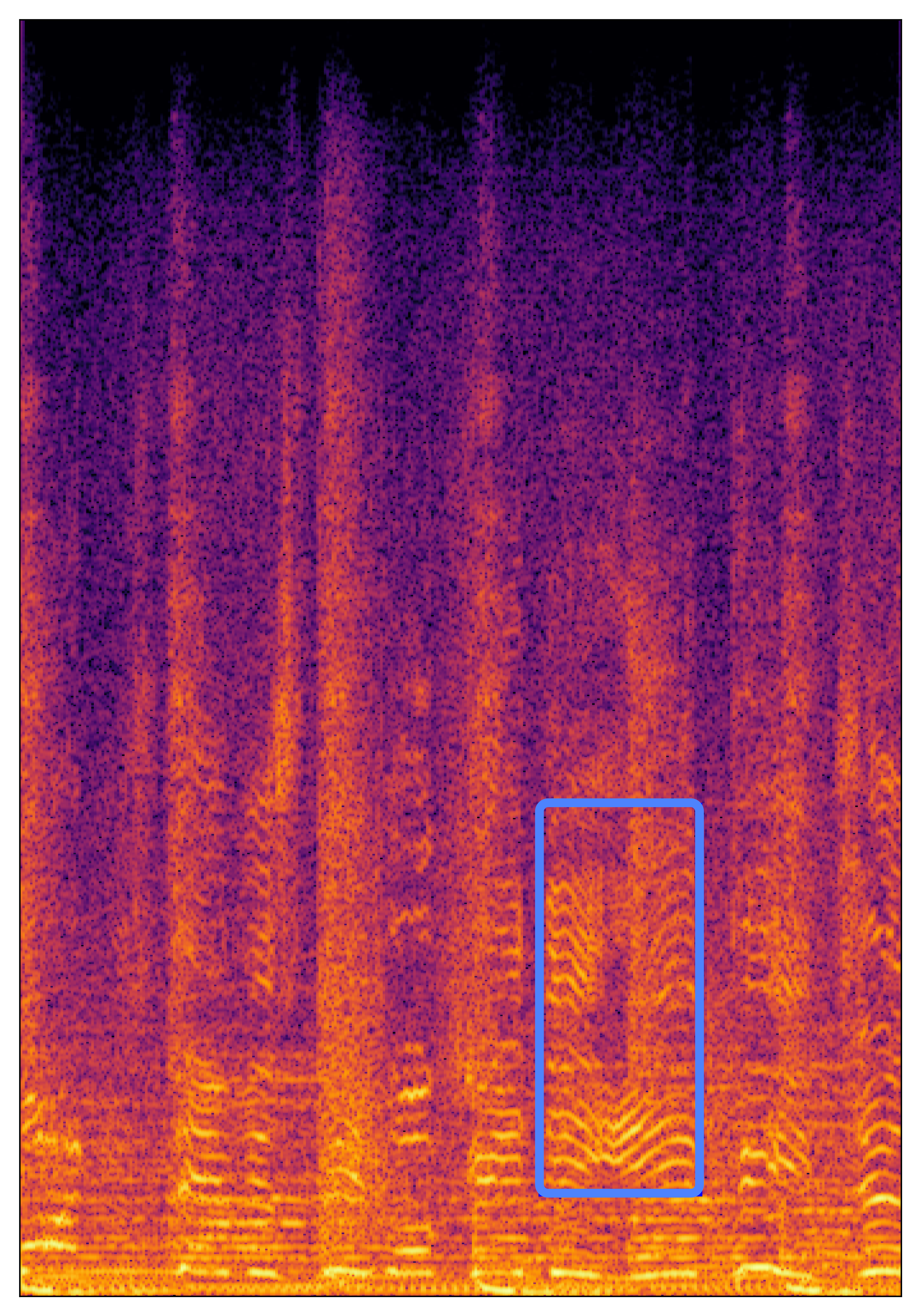}%
        \label{fig:sub-a}%
    }
    \hfill
    % 第二个子图
    \subfloat[GT (MP3 Compressed)]{%
        \includegraphics[width=0.23\textwidth]{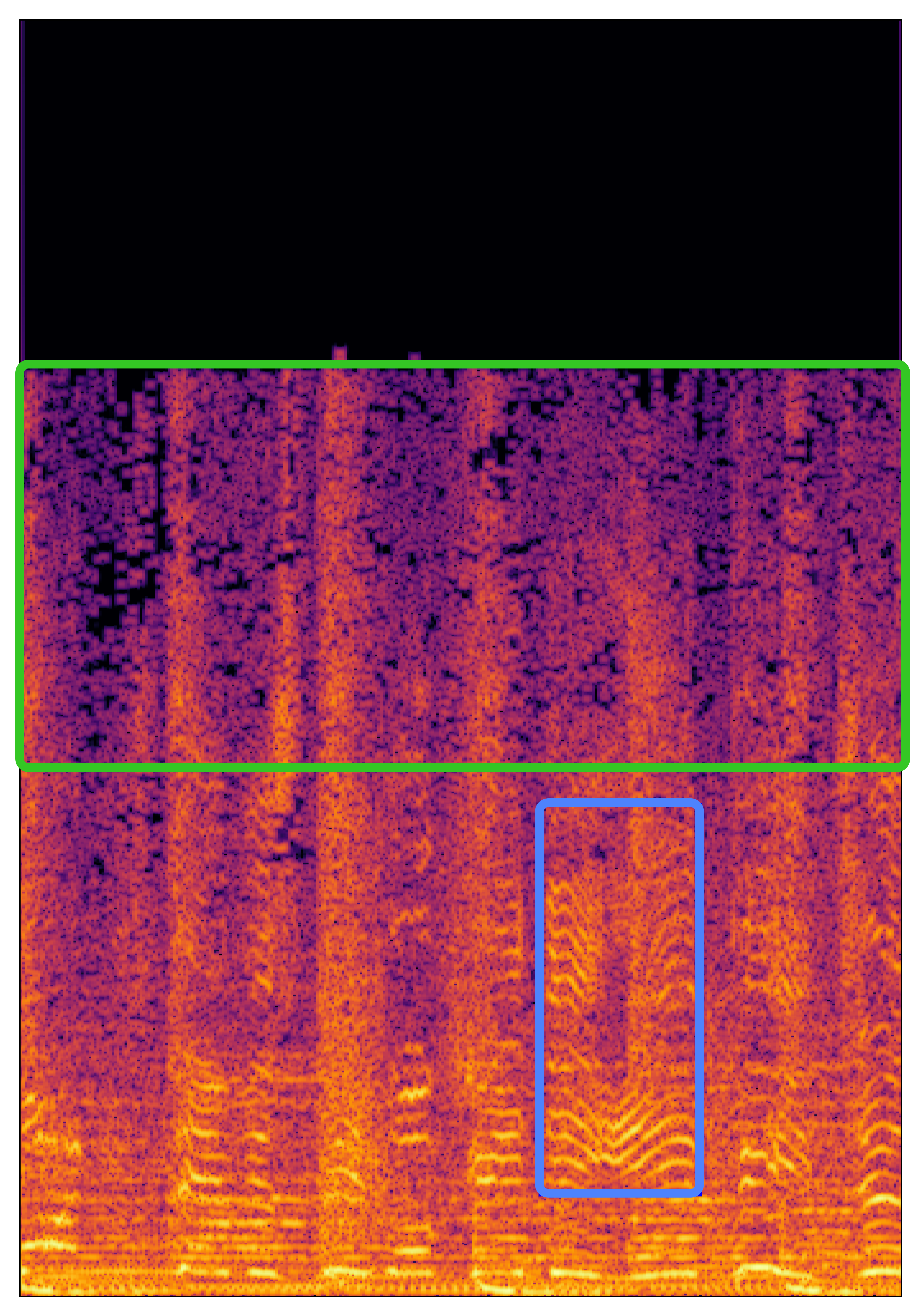}%
        \label{fig:sub-b}%
    }
    \hfill
    % 第三个子图
    \subfloat[Proposed VAE]{%
        \includegraphics[width=0.23\textwidth]{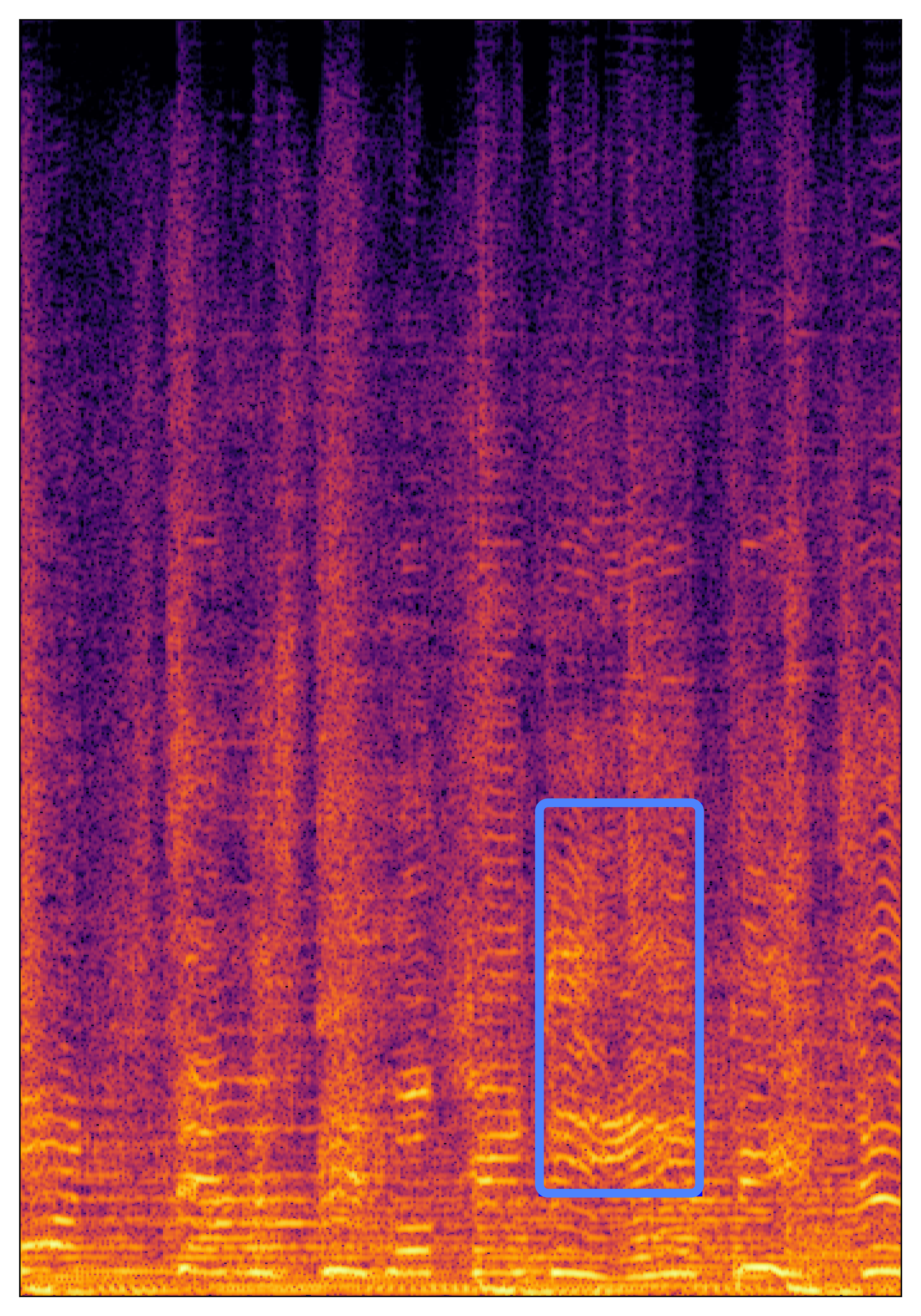}%
        \label{fig:sub-c}%
    }
    \hfill
    % 第四个子图
    \subfloat[Stable Audio 2 VAE]{%
        \includegraphics[width=0.23\textwidth]{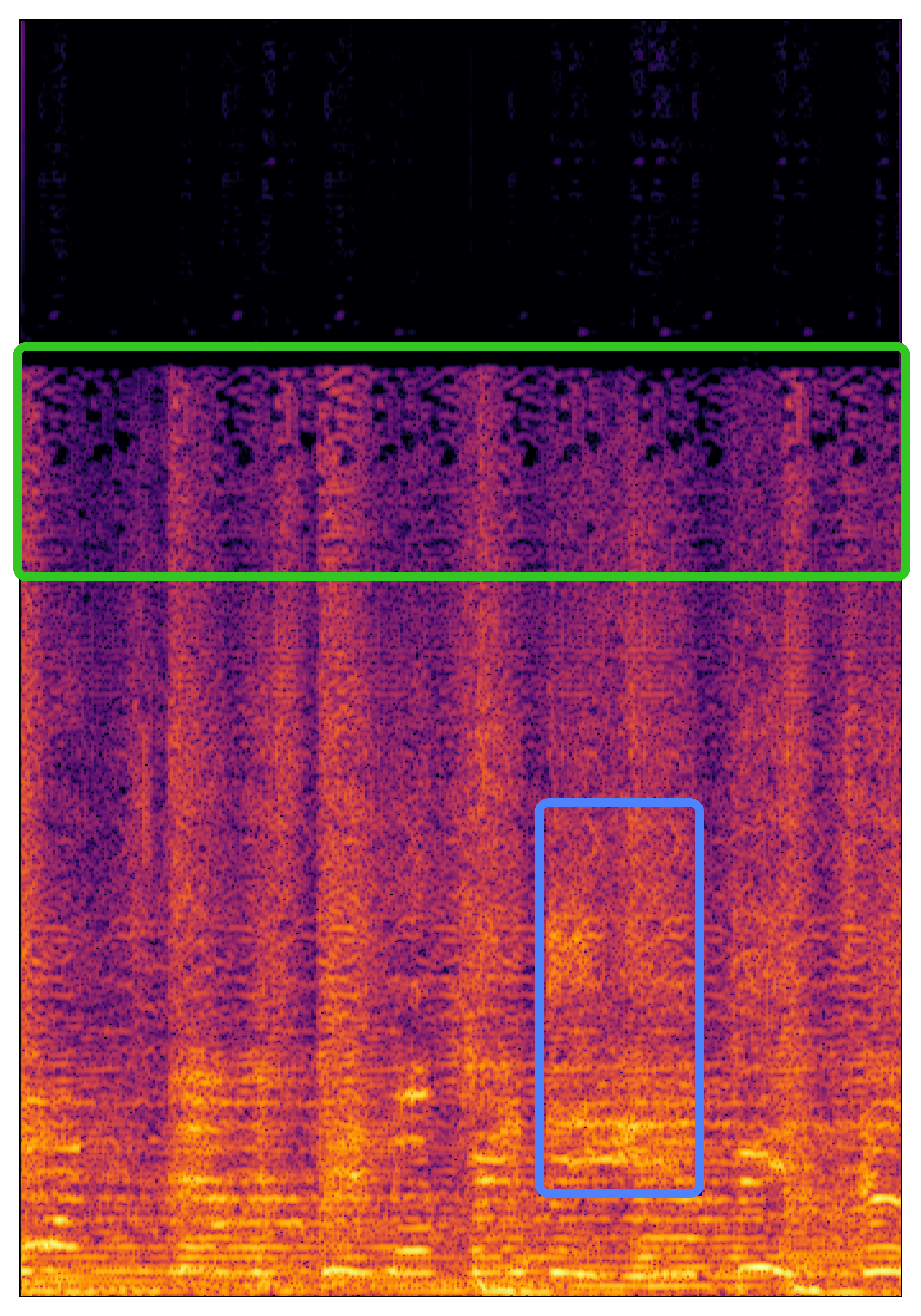}%
        \label{fig:sub-d}%
    }
    % 第五个子图
    % \subfloat[Music2Latent]{%
    %     \includegraphics[width=0.23\textwidth]{figs/1308818967_2_music2latent.png}%
    %     \label{fig:sub-e}%
    % }
    % % 第六个子图
    % \subfloat[AudioGen]{%
    %     \includegraphics[width=0.23\textwidth]{figs/1308818967_2_agc.png}%
    %     \label{fig:sub-f}%
    % }
    \caption{Visualization of Spectrograms from (a) lossless ground-truth, (b) ground-truth after MP3 compression, (c) MP3 reconstructed by proposed VAE, (d) MP3 reconstructed by Stable Audio VAE. Boxed regions indicate areas to be analyzed in the main text.}
    \label{fig:spec}
%    \vspace{-8pt}
\end{figure}

\begin{table*}[ht]
\centering
\renewcommand\arraystretch{1.2}
\caption{Objective and subjective evaluation results of comparison and ablation systems for song generation. DiffRhythm-base and DiffRhythm-full represent DiffRhythm with generation length of 1m35s and 4m45s respectively, and w/o align stands for the ablation system without sentence-level alignment.}
\label{tab:song}
\resizebox{1.0\linewidth}{!}{
\begin{tabular}{lccccccl}
\hline
                       & PER$\downarrow$ & FAD$\downarrow$ & Musicality$\uparrow$ & Quality$\uparrow$ & Intelligibility$\uparrow$ & Generation Length & RTF$\downarrow$   \\ \hline
GT (VAE-reconstructed) & 16.14\%         & 0.88            & 4.68$\pm0.06$        & 4.43$\pm0.06$     & 4.17$\pm0.03$             & -                 & -     \\
SongLM                 & 21.35\%         & 1.92            & 4.27$\pm0.04$        & 4.06$\pm0.03$     & 3.44$\pm0.03$             & 120 s             & 1.717     \\
DiffRhythm-base        & 17.47\%         & 2.11            & 4.14$\pm0.07$        & 4.19$\pm0.05$     & 3.80$\pm0.04$             & 95 s              & 0.037 \\
DiffRhythm-full        & 18.02\%         & 2.25            & 4.02$\pm0.02$        & 4.21$\pm0.04$     & 3.68$\pm0.07$             & 285 s             & 0.034 \\
\hspace{1em}w/o align & -               & 3.16            & 4.07$\pm0.05$        & 3.04$\pm0.02$     & -                         & 95 s              & 0.037 \\ \hline
\end{tabular}}
%\vspace{-8pt}
\end{table*}

\subsection{Evaluation Metrics}
\textbf{Objective Evaluation}\quad
To evaluate the quality of waveform reconstruction, we calculate STOI~\cite{stoi}, PESQ~\cite{pesq} and Mel cepstral distortion (MCD)~\cite{mcd}. For evaluating the quality song generation, we utilize the Phoneme Error Rate (PER) and Fréchet Audio Distance (FAD)~\cite{fad}. We employ FireRedASR~\cite{fireredasr}, which is currently the state-of-the-art Automatic Speech Recognition (ASR) model, to recognize the vocal content of the generated songs. FireRedASR not only achieves remarkably high performance for vocals but is also robust in recognizing singing vocals. Given that ASR may perceive vocal content as different words with consistent pronunciation, such errors do not accurately reflect actual vocal intelligibility; therefore, we calculate the PER instead of the Word Error Rate (WER) or Character Error Rate (CER). Realtime factor (RTF) is also calculated using Nvidia RTX 4090 to demostrate the computational efficiency of the comparison models.

\textbf{Subjective Evaluation}\quad
We conducted mean opinion score (MOS) listening tests for subjectively evaluation. Specifically, 30 listeners participated in rating each generated song sample on a scale from 1 to 5 across three aspects: musicality, quality and intelligibility. 

\section{Evaluation Results}

\subsection{Waveform Reconstruction}
%We compare the performance in speech reconstruction of the proposed VAE with Music2Latent~\cite{music2latent} and Stable Audio 2~\cite{stable-audio}. We tested two sets of metrics, one with lossless audios as model inputs and the target for metric computation, and one with lossy audio compressed with MP3 as input while the target for metric computation is still lossless audios. For the second test, only the proposed VAE was tested, considering that the two comparison models have no restoration capability. The results are shown in Table~\ref{tab:ae}. With lossless input, all the metrics outperform the comparison models, demonstrating the excellent refactoring capabilities of our proposed VAE. Using lossy audio as input, the metrics of the proposed VAE show only a slight degradation compared to the lossless metrics, proving the excellent spectral restoration capability of our proposed VAE

%Furthermore, we visualized the spectrograms of ground-truth and reconstructed songs by comparison systems. Figure~\ref{fig:spec} shows that, after MP3 compression, audio loses high frequencies and can be hollow in the low frequencies. Taking the compressed MP3 as input, the proposed VAE not only fills the missing high frequencies, but also filling in the hollows in the low frequencies (green box). The audio reconstructed by the Stable Audio 2 VAE retains the compression loss of the MP3. On the other hand, as shown in the red boxed part of the figure, our proposed VAE has significantly better reconstruction ability for vocal harmonics than the open-source model, generating clearer vocals.

We conduct a comprehensive evaluation of waveform reconstruction performance comparing our VAE with two popular open-sourced baselines: Music2Latent~\cite{music2latent} and Stable Audio 2~\cite{stable-audio}. The evaluation protocol consists of two experimental settings: (1) lossless-to-lossless reconstruction using lossless audio inputs, and (2) lossy-to-lossless reconstruction using MP3-compressed inputs while maintaining lossless reference targets. As shown in Table~\ref{tab:ae}, the proposed method achieves superior performance across all metrics in both experimental conditions. Specifically, under lossless input conditions, our model demonstrates 3.8\% and 12.3\% relative improvements in STOI and PESQ respectively over the best baseline, while maintaining comparable MCD scores. More importantly, when processing lossy MP3 inputs - a scenario where baseline models completely fail due to their lack of restoration capability - our method maintains robust performance with only minimal degradation on all three metrics compared to the lossless condition.

To further validate the reconstruction quality, we perform spectral visualization comparing the proposed VAE with baseline models. Figure~\ref{fig:spec} reveals three key observations: First, MP3 compression artifacts manifest as both high-frequency attenuation (above 32 kHz) and mid-frequency hollowing effects (16 kHz - 32 kHz). Second, our VAE successfully addresses both artifact types - it not only generates missing high-frequency components but also restores the spectral continuity in mid-frequency regions (green box). Third, the proposed model demonstrates superior harmonic reconstruction capability for vocal components, particularly in preserving formant structures, resulting in significantly clearer vocal components compared to the open-source baseline that produces vague harmonics (blue box).

\subsection{Song Generation}
%In this paper, we present DiffRhythm, the first full-diffusion-based model for song generation capable of producing high-quality 44.1kHz stereo songs up to 4m45s. Both the model architecture and data processing pipeline are embarassingly simple, demonstrating excellent scalability potential. Leveraging flow-matching objectives, our framework achieves efficient sampling within 20 steps, enabling generation of 4-minute singing vocals in under 30 seconds. Extensive experimental results validate the effectiveness of our approach. showing the strong song generation capability of DiffRhythm. Code and model weights are fully open-sourced.
For the evaluation of song generation, we compare DiffRhythm with SongLM~\cite{songeditor}, the samples of SongLM were kindly provided by the authors.
As shown in Table~\ref{tab:song}, the GT songs reconstructed via VAE naturally achieves the best performance across all metrics, serving as an upper bound for synthesized song quality. Compared to the SongLM baseline, DiffRhythm models achieve superior quality and intelligibility while maintaining comparable musicality. The significant 18.2\% relative reduction in PER further confirms our model's improved vocal content clarity. However, SongLM shows slightly better FAD and musicality scores, suggesting room for improvement in long-term acoustic consistency and melodic expression.

The full-length DiffRhythm variant exhibits marginally degraded PER and FAD than its base version, likely due to increased modeling complexity for longer sequences. Notably, both variants maintain RTF below 0.04, achieving a $\sim50\times$ speedup over SongLM, highlighting the  computational efficiency of our diffusion-based approach compared to autoregressive language models.

Our ablation study reveals the critical role of sentence-level alignment. As shown in Table~\ref{tab:song}, removing this approach catastrophically degrades intelligibility (unmeasurable PER and intelligibility MOS) and audio quality, though interestingly preserves basic musical structure. This validates our hypothesis that sentence-level alignment is essential for establishing semantic correspondence between tight lyrics and vocals.

The relatively high PER across all systems may stem from using mixed audio containing both vocal and accompaniment without source separation for ASR evaluation, as accompaniment likely interferes with ASR recognition.

\section{Conclusion}

In this paper, we propose DiffRhythm, the first full-diffusion-based model capable of generating complete stereo songs of 4m45s in just 10 seconds, featuring both vocals and accompaniment. The model's elegant design eliminates the need for complex multi-stage cascading modeling and laborious data preprocessing, facilitating scalability. DiffRhythm’s non-autoregressive structure ensures rapid inference speeds while preserving high musical quality and lyrical intelligibility. Extensive experimental results demonstrate the effectiveness of our approach and underscore the robust song generation capabilities of DiffRhythm. Furthermore, the system's simplicity and open accessibility—through our release of code and pre-trained models—establish a new foundation for scalable, end-to-end research in song generation.

\section{Limitations}
While DiffRhythm demonstrates good capability to generate high-quality full-length songs, two important aspects remain unexplored in our current framework. First, the functionality for editing specific segments within generated compositions has not been investigated. Incorporating random masking of latent representations during training could enable song editing (inpainting) and continuation (outpainting). %Second, although DiffRhythm employs a short audio clip as a style prompt, achieving finer-grained control through natural language descriptions would significantly enhance its style controllability.
Second, the model employs short audio clips as style references, integrating natural language conditioning mechanisms would enable finer-grained stylistic control through textual descriptions. This improves the flexibility of the model by eliminating the need for audio references.
\bibliography{custom}

\appendix

% \section{Example Appendix}
% \label{sec:appendix}

% This is an appendix.

\end{document}